\documentclass[fleqn,usenatbib]{mnras}

\usepackage{newtxtext,newtxmath}

\usepackage[T1]{fontenc}
\usepackage{ae,aecompl}

\usepackage[dvipsnames]{xcolor}

\usepackage{graphicx}	
\usepackage{amsmath}	
\usepackage{amssymb}	

\title[Hills and holes in microlensing light curve]{Hills and holes in the microlensing light curve due to plasma environment around gravitational lens}

\author[O. Yu. Tsupko and G. S. Bisnovatyi-Kogan]{
Oleg Yu. Tsupko$^{1}$\thanks{E-mail: tsupko@iki.rssi.ru, tsupkooleg@gmail.com (OYuT); ORCID iD: https://orcid.org/0000-0002-2159-8350}
and Gennady S. Bisnovatyi-Kogan$^{1,2,3}\thanks{E-mail: gkogan@iki.rssi.ru (GSBK); ORCID iD: https://orcid.org/0000-0002-2981-664X}$
\\
$^{1}$Space Research Institute of Russian Academy of Sciences, Profsoyuznaya 84/32, Moscow 117997, Russia\\
$^{2}$National Research Nuclear University MEPhI (Moscow Engineering Physics Institute), Kashirskoe Shosse 31,\\ \qquad Moscow 115409, Russia\\
$^{3}$Moscow Institute of Physics and Technology, 9 Institutskiy per., Dolgoprudny, Moscow Region, 141701, Russia
}

\date{Accepted XXX. Received YYY; in original form ZZZ}

\pubyear{2015}

\begin{document}
\label{firstpage}
\pagerange{\pageref{firstpage}--\pageref{lastpage}}
\maketitle

\begin{abstract}
In this paper, we investigate the influence of the plasma surrounding the gravitational lens on the effect of microlensing. In presence of plasma around the lens, the deflection angle is determined by both the gravitational field of the lens and the chromatic refraction in the inhomogeneous plasma. We calculate microlensing light curves numerically for point-mass lens surrounded by power-law density distribution of plasma. A variety of possible curves is revealed, depending on the plasma density and frequency of observations. In the case of significant influence of plasma, the shape of microlensing light curve is strongly deformed in comparison with vacuum case. If the refractive deflection is large enough to compensate or to overcome the gravitational deflection, microlensing images can completely disappear for the observer. In this case, the remarkable effect occurs: formation of a 'hole' instead of a 'hill' in the center of microlensing light curve. Observational prospects of 'hill-hole' effect in different microlensing scenarios are discussed.
\end{abstract}

\begin{keywords}
gravitational lensing: micro -- plasmas
\end{keywords}



\section{Introduction}

The theory of gravitational lensing deals usually with a propagation of light rays in vacuum. In vacuum, ray paths and deflection angles are independent on the photon frequency, and therefore all effects of gravitational lensing are achromatic (in the frame of geometrical optics). In particular, magnification curves for microlensing events are supposed to look the same at all wavelengths. Meanwhile, cosmic space is filled with plasma and plasma is also expected to be concentrated around compact objects. Under these conditions, a question arises: how does the plasma influence the phenomena of gravitational lensing?

In this article, we examine the effect of plasma on the microlensing phenomenon (for general discussion of microlensing see, e.g., \cite{GL1, GL2, Mao-review, Dodelson-GL, Keeton-book}). Presence of plasma leads to change of the deflection angle due to the effect of refraction. Moreover, since plasma is a dispersive medium, lensing becomes chromatic. Positions of images, magnification factors and, what is important for this article, microlensing light curves depend on the frequency at which observations are made.

Magnitude of plasma influence is determined by the ratio of the plasma frequency and the photon frequency. Lower photon frequency (longer wavelength) or higher plasma density leads to larger refractive deflection. We calculate the total deflection of the ray as a sum of the gravitational deflection in vacuum and chromatic refraction in an inhomogeneous plasma, using the approximation that both angles are small. Important feature of our analysis is that the plasma deflection is not assumed to be much smaller than the gravitational one.

When a light ray travels through the plasma atmosphere of a star or through the accretion disk around a black hole, the influence of the plasma on the light deflection can be significant due to the high concentration of the plasma.
If the plasma deflection is large enough, it can fully compensate or overcome the gravitational deflection. In this paper, we show that the microlensing images can disappear (become invisible) because light rays from source will be diverged by the lens and will not reach the observer. This leads to the remarkable effect: formation of a 'hole' instead of a 'hill' in the center of microlensing light curve. We calculate the microlensing light curves numerically for point-mass lens surrounded by power-law density distribution of plasma. A variety of possible curves is revealed, depending on the plasma density and frequency of observations. Observational prospects depend on the lens mass, the plasma distribution and the frequency of observations involved. We discuss possible applications of 'hill-hole' effect to different microlensing scenarios.

The paper is organized as follows. In the next section we present the overview of previous results in research of plasma effects in gravitational lensing. In Section \ref{sec-angle} we discuss the properties of overall deflection angle and approximations used. In Section \ref{sec-le} we introduce the lens equation in presence of both gravitational and refractive plasma deflections. In Section \ref{sec-num-magnif} we describe our method for numerical calculation of microlensing light curve in presence of plasma. Sections \ref{sec-k3} and \ref{sec-k32} are devoted to results of numerical calculations for particular plasma density profiles. In Section \ref{sec-observ} we introduce the physical criterium characterizing the possibility of 'hole' formation and discuss the observational prospects for quasar microlensing and Galactic microlensing. Section \ref{sec-Th-FF} is devoted to other physical effects in a dense plasma. Section \ref{sec-conclusions} is Conclusions.

\section{Gravitational lensing in plasma: state of research}

For gravitational lensing, the main interest is the change in the deflection angle due to the presence of plasma. In the simplest case, when angles are small, we can separate gravitational and plasma deflections from each other, assuming that gravitational deflection is the same as in vacuum and plasma deflection is refractive deflection due to medium inhomogeneity. Under these conditions, to calculate the gravitational deflection angle, it is sufficient to use the linearized GR (which leads to the Einstein angle formula), and for refraction, we can assume that the refractive index of plasma is close to unity. In this approximation, the overall deflection angle was calculated in papers of \citet{Muhleman-1966, Muhleman-1970}, in application to the deflection of radio signals in the solar corona plasma. The same approximation was subsequently used in book of \citet{Bliokh-Minakov-1989} who discussed  different problems in context of gravitational lensing for the first time. In particular, the authors qualitatively discussed the number and the positions of the images.

In the approach described above, the deflections of light due to gravity and due to refraction in a non-homogeneous plasma were considered as small, and independent of each other effects. There are situations and problems when this approach is no longer enough.
It happens, when
the light travels close enough to the black hole, so that the light bending is not small. In addition, even  a plasma with uniform density distribution can affect the magnitude of the gravitational deflection itself \citep{BK-Tsupko-2009}.

The deflection angle is generally determined by a complex combination of effects of gravity, refraction and dispersion. More general consideration in general relativity of a light propagation in presence of medium was done by \citet{Synge-1960}, and subsequently investigated by \citet{Bicak-1975} who first applied it for plasma, see also \citet{Kulsrud-Loeb-1992}. Part of the monograph of \citet{Perlick-2000} is devoted to investigation of light ray propagation in presence of both gravity and plasma. In particular, the exact expression for the deflection angle for the equatorial plane of the Kerr black hole was obtained in the form of integral.

Extensive study of gravitational lensing in presence of plasma has been started only about one decade ago. On basis of Synge's approach, the deflection angles in weak deflection case were derived, for light rays propagating in Schwarzschild metric in presence of non-homogeneous plasma \citep{BK-Tsupko-2009, BK-Tsupko-2010}, see also \citet{BK-Tsupko-2015}. In particular, it has been shown that in a homogeneous plasma, where refractive deflection is absent, the gravitational deflection itself differs from vacuum case and depends on the photon frequency due to dispersive properties of plasma. For the case of homogeneous plasma, positions and magnifications of images were calculated analytically \citep{BK-Tsupko-2010}. \citet{Er-Mao-2014} have continued studies of gravitational lensing in plasma in case of weak deflection taking into account vacuum gravitational deflection and refractive deflection in non-homogeneous plasma. They have performed a numerical modelling of a strong lens system and have shown that changes in the angular position of images due to refraction in plasma are possible to be detected in the low frequency radio observations.

In the case when a compact object, like a black hole or a neutron star, is surrounded by plasma, the rays of the light can experience a strong deflection. In the case of a gravitational lensing by a black hole, the rays can make one or more revolutions around it before reaching the observer. In presence of plasma, the properties of high-order images (also called relativistic \citep{Virbhadra-2000, Bozza-2001, Perlick-review}) formed by such rays have been analytically studied in the article of \citet{Tsupko-BK-2013}. A complex behavior of light rays near compact gravitating objects surrounded by inhomogeneous plasma was investigated in a series of works of \citet{Rogers-2015, Rogers-2016, Rogers-2017a, Rogers-2017b}. Curved ray paths were calculated numerically for various spherically symmetric power-law plasma distributions. Strong bending of light rays takes place also in the case of a shadow of black hole surrounded by plasma, see \citet{Perlick-Tsupko-BK-2015}, \citet{Perlick-Tsupko-2017}, \citet{Huang-2018}, \citet{Yan-2019}, see also \citet{Gralla-2019}, \citet{Johnson-2019}, \citet{Narayan-2019}.

For investigation of gravitational lensing in plasma in Kerr space-time we refer to \citet{Perlick-2000}, \citet{Morozova-2013}, \citet{Perlick-Tsupko-2017}, \citet{Liu-2017}, \citet{Kimpson-2019a, Kimpson-2019b}. For the wave effects in presence of Solar gravity and Solar corona, see \citet{Turyshev-2019a, Turyshev-2019b}. For some recent studies see \citet{Perlick-Schwarz-2017, Crisnejo-2018, Crisnejo-Rogers-2019, Er-Rogers-2018, Er-Rogers-2019a, Er-Rogers-2019b, Ovgun-2019, Bratislava-2019, Crisnejo-Jusufi-2019}. Recent review of plasma effects in gravitational lensing was made by \citet{BKTs-Universe-2017}.

\section{Overall deflection angle} \label{sec-angle}

Let us consider a photon flying past a spherical or point massive body of a mass $M$ surrounded by a spherically symmetric plasma distribution with a number density $N(r)$. Assuming that the deflection angle of the photon is small, it can be written as the following function of the photon impact parameter $b$ (see, e.g., \citet{BK-Tsupko-2015}):
\begin{equation} \label{total-deflection-b}
\hat{\alpha}(b) = \alpha_{grav}(b) + \alpha_{refr}(b) \, .
\end{equation}
Here the first term is the gravitational deflection in vacuum:
\begin{equation} \label{einst}
\alpha_{grav}(b) = \frac{4m}{b} \, ,
\end{equation}
where $m=GM/c^2$ is the mass parameter. The second term in (\ref{total-deflection-b}) is the non-relativistic refractive deflection.

Let us consider light rays in non-magnetized cold electron plasma with the refractive index
\begin{equation} \label{n}
n^2  =  1 - \frac{\omega_p^2}{\omega^2} \, ,
\end{equation}
where
\begin{equation} \label{K-e}
\omega_p^2 = K_e N, \quad K_e  = \frac{4\pi e^2}{m_e}  \, .
\end{equation}
Here $N$ is the number density of the electrons in the plasma, $\omega_p$ is the electron plasma frequency, $\omega$ is the photon frequency, $e$ is the electron charge, $m_e$ is the electron mass.

If the unperturbed path of the photon is parallel to the $z$-axis, deflection angle can be written as \citep{BK-Tsupko-2009, BK-Tsupko-2010, BK-Tsupko-2015}:
\begin{equation} \label{alpha-refr-b}
\alpha_{refr}(b) =  \frac{K_e}{\omega_0^2}   \int \limits_0^\infty \frac{\partial N}{\partial b} \, dz .
\end{equation}
Here $\omega_0$ is the photon frequency at infinity. Since we neglect the gravity in calculation of the refraction, there is no difference between frequency at infinity and at some other observer's position, so $\omega_0$ equals to $\omega$ in formula (\ref{n}), see discussion below.

Properties of the overall deflection angle:

(i) In this paper, we consider gravitational deflection and refractive deflection as separate effects, and the overall deflection angle is calculated as a simple summation of two separate deflections. In this approximation, gravity does not influence the photon frequency in eq.(\ref{n}). If we would like to go beyond such approximate approach (to consider a mutual influence of gravity and plasma, strong bending situations, black hole shadow etc), the influence of gravity on the photon frequency should be taken into account. In the rigorous approach (see \citet{Synge-1960, BK-Tsupko-2010, Perlick-Tsupko-BK-2015}), the photon frequency $\omega$ in (\ref{n}) is a frequency measured by a static observer at the position $r$, and it is a function of the space coordinates due to gravitational redshift. For example, for the Schwarzschild gravity, $\omega(r)$ is connected with $\omega_0$ as
\begin{equation} \label{redshift}
\omega(r) = \frac{\omega_0}{\sqrt{1-2m/r}}  \, .
\end{equation}
In the exact calculations of deflection, the frequency $\omega(r)$ should be used in (\ref{n}) from the very beginning in all calculations. In our paper we use approximate approach, and in the refraction term we assume that $\omega(r) \simeq \omega_0$.

(ii) We also neglect here the plasma influence on the gravitational deflection itself, assuming that gravitational deflection is the same as in vacuum. Namely, we neglect corrections to the gravitational deflection associated with the presence of plasma derived in \citet{BK-Tsupko-2009, BK-Tsupko-2010}. These corrections should not be confused with non-relativistic refractive terms. As in the previous comment, for calculation of these corrections, the refractive index (\ref{n}) with (\ref{redshift}) should be used.

(iii) In eq.(\ref{alpha-refr-b}), we have $N=N(r)$ and $r=\sqrt{b^2+z^2}$, so the expression under the integral sign is a function of $b$ and $z$. To calculate the deflection angle, we perform partial differentiation with respect to $b$ at constant $z$, and then, after integration along $z$ at constant $b$, obtain deflection angle as the function of $b$.

(iv) Vacuum gravitational deflection is usually considered as positive, see (\ref{einst}). The refractive deflection (\ref{alpha-refr-b}) can be both positive or negative, depending on the density profile. Usually the density of plasma in different models decreases with radius ($dN/dr<0$), therefore the refractive deflection is usually opposite to the vacuum gravitational deflection: ($\alpha_{refr}(b)<0$).

(v) In the case of inhomogeneous plasma with a power-law number density
\begin{equation} \label{N-power}
N(r) = N_0 \left( \frac{r_0}{r} \right)^k ,
\end{equation}
\begin{equation}
N_0 = \mbox{const}, \; r_0 = \mbox{const}, \; k=\mbox{const} > 1 \, ,
\end{equation}
the refractive deflection is \citep{Bliokh-Minakov-1989, BK-Tsupko-2010, BK-Tsupko-2015}
\begin{equation} \label{angle-refr-power}
\alpha_{refr}(b) = - \frac{K_e}{\omega_0^2} N_0 \left(\frac{r_0}{b}\right)^k \frac{\sqrt{\pi} \,   \Gamma\left(\frac{k}{2} + \frac{1}{2}\right)}{\Gamma\left(\frac{k}{2}\right)} \, ,
\end{equation}
\begin{equation}
\Gamma(x) = \int \limits_0^\infty t^{x-1} e^{-t} dt \,  .
\end{equation}

\section{Lens equation in presence of gravity and plasma} \label{sec-le}

\begin{figure*}
\begin{center}
\includegraphics[width=0.95\textwidth]{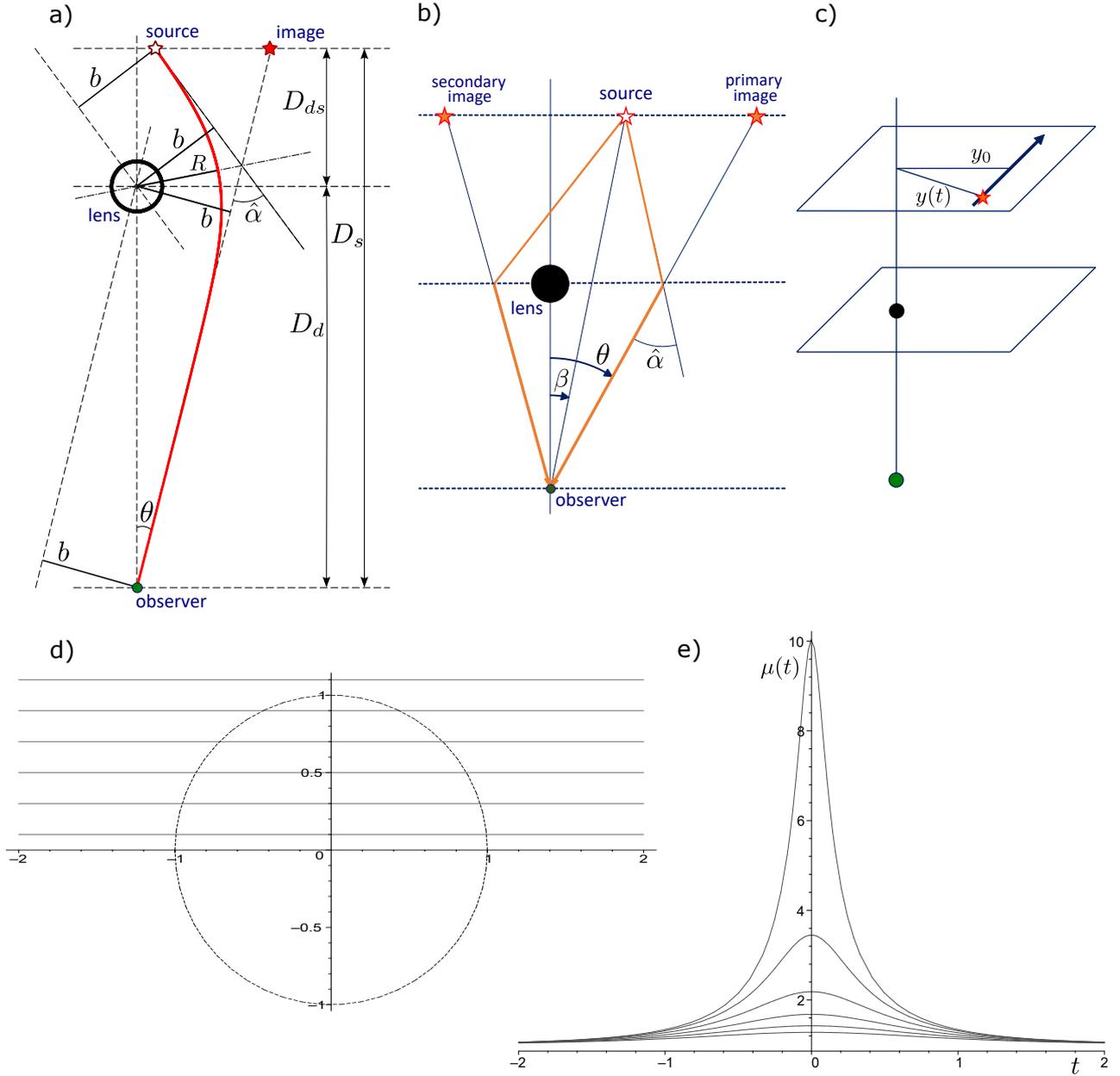}
\end{center}
\caption{Effect of microlensing in vacuum. a) An example of the trajectory of a light ray calculated in vacuum. The impact parameter of the light ray $b$ and the distance of the closest approach $R$ are different values, but they are supposed to be equal in weak deflection approximation. b) Geometry of a point-mass lensing and notations in the lens equation. Two images are formed in the case of lensing in vacuum. c) Microlensing event: a source moves with the impact parameter $y_0$ (in the normalized units) relative to the observer-lens axis. The value $y_0$ should not be confused with the impact parameter $b$ of the light ray. The instantaneous position of the source is given by $y(t)$, minimum of $y(t)$ equals to $y_0$. d) Six trajectories of the source with different impact parameters, namely, $y_0 = 0.1, \; 0.3,\; 0.5,\; 0.7,\; 0.9,\; 1.1$. The dashed ring of unit radius corresponds to Einstein ring. e) Corresponding microlensing light curves $\mu(t)$ for impact parameters of the source which are shown in d). Higher curve corresponds to smaller impact parameter. See also Fig.2, page 458 of \citet{GL2}, and the interactive tool for plotting microlensing light curves in the website 'microlensing-source.org'.}
\label{fig-microlens}
\end{figure*}

For the axially symmetric mass distribution around a lens, the lens equation is one-dimensional. It connects the angular position $\beta$ of a source and angular positions $\theta$ of the lensed images through the deflection angle $\hat{\alpha}$, see Fig. \ref{fig-microlens}a,b. For a general case of arbitrary positions of source, lens and observer and arbitrary magnitude of deflection angle, the lens equation can be found in \citet{Frittelli-2000}, \citet{Perlick-2004b}, \citet{Bozza-2008b}.

In most cases related with applications, the simplifications are used. First of all, all angles are supposed to be small: $\beta, \theta, \hat{\alpha} \ll 1$. The second, the distances between the source, lens, and observer are assumed to be much larger than the size of the lens (thin lens approximation).

Angles $\beta$ and $\theta$ are calculated from the line connecting the lens with an observer, see Fig. \ref{fig-microlens}a,b. Usual convention is the following: $\beta$ is supposed to be positive, and $\theta$ can be both positive (image is on the same side from lens as a source) and negative (image is on the opposite side), see Fig. \ref{fig-microlens}b. The case $\beta=0$ corresponds to perfect alignment of source, lens and observer; in this case the Einstein ring can be formed. Photon impact parameter $b$ and $\theta$ are related by formula $b=D_d |\theta|$.

In approximations described above, the lens equation has a simple form:
\begin{equation} \label{le}
\beta = \theta - \tilde{\alpha}(|\theta|) \frac{\theta}{|\theta|} \, ,
\end{equation}
where
\begin{equation}
\tilde{\alpha}(|\theta|) = \frac{D_{ds}}{D_s} \hat{\alpha}(D_d |\theta|) \, .
\end{equation}

Note that very often, the lens equation for axially symmetric case is written as $\beta = \theta - \tilde{\alpha}(\theta)$, see, e.g., eq.(35), p.31 in \citet{GL2}. In such cases, it is assumed that the deflection angle has the property $\tilde{\alpha}(-\theta) = - \tilde{\alpha}(\theta)$, see eq.(36), p.32 in \citet{GL2}. In our case, the argument in the angle (\ref{total-deflection-b}) has been supposed to be positive. Therefore, in order to correctly deal with the negative values of variable $\theta$, the lens equation is written as (\ref{le}), see eq. (17) of \citet{Suyu-2012} and eq. (13), p.104 of \citet{GL2}, see also eq.(2.80) in \citet{Keeton-book}.

In order to simplify eq.(\ref{le}), let us introduce widely used normalized variables for the source and the image positions,
\begin{equation}
y = \frac{\beta}{\theta_E} \, , \quad x = \frac{\theta}{\theta_E} \, .
\end{equation}
Here $\theta_E$ is the vacuum Einstein angular radius
\begin{equation}
\theta_E = \sqrt{ \frac{4m D_{ds}}{D_d D_s} }  \,  ,
\end{equation}
which corresponds to the angular size of the Einstein ring observed in the case of a perfect alignment (for the point-mass lens).

We obtain in new variables:
\begin{equation} \label{le-final}
y = x - \alpha(|x|) \frac{x}{|x|} \,  ,
\end{equation}
with a normalized total deflection angle
\begin{equation}
\alpha(|x|) = \frac{1}{|x|} + \frac{1}{\theta_E} \frac{D_{ds}}{D_s} \alpha_{refr}(D_d \theta_E |x|) \,  .
\label{alpha}
\end{equation}
Here, the first term is simplified presentation for a gravitational deflection, the second term is connected with a refractive deflection. Combination $D_d \theta_E |x|$ is an argument of $\alpha_{refr}$.

Below we consider only cases of the falling density profiles, for which $\alpha_{refr}<0$. Therefore, for convenience, we introduce a positive function
\begin{equation} \label{B-defin}
B(|x|) \equiv - \frac{1}{\theta_E} \frac{D_{ds}}{D_s} \alpha_{refr}(D_d \theta_E |x|) \,  .
\end{equation}
By physical meaning, the function $B(x)$ is a refractive deflection angle taken with an opposite sign (normalized by distances ratio and by Einstein angular radius). This function is positive for spherically symmetric profiles with a density decreasing with radial coordinate, and negative in the opposite case.

Using $B(|x|)$, the deflection becomes
\begin{equation} \label{norm-total-delf}
\alpha(|x|) = \frac{1}{|x|} - B(|x|) \, .
\end{equation}
In the particular case of a power-law distribution (\ref{N-power}) leading to the refractive deflection angle (\ref{angle-refr-power}), the function $B(|x|)$ can be represented as:
\begin{equation} \label{B-power}
B(|x|) = \frac{B_0}{|x|^k}  \, ,
\end{equation}
where we have introduced a constant
\begin{equation} \label{B0}
B_0 = \frac{1}{\theta_E} \frac{D_{ds}}{D_s} \frac{K_e}{\omega_0^2} N_0 \left(\frac{r_0}{D_d \theta_E}\right)^k \frac{\sqrt{\pi} \,   \Gamma\left(\frac{k}{2} + \frac{1}{2}\right)}{\Gamma\left(\frac{k}{2}\right)} \, .
\end{equation}
The constant $B_0$ contains both values connected with a plasma and with a gravity, and characterizes the magnitude of the plasma influence in comparison with the gravitational one. Using $\omega_p^2(r) = K_e N(r)$, it can be also rewritten as
\begin{equation} \label{B0-2}
B_0 = \frac{1}{\theta_E} \frac{D_{ds}}{D_s} \frac{\omega_p^2(D_d \theta_E)}{\omega_0^2} \frac{\sqrt{\pi} \,   \Gamma\left(\frac{k}{2} + \frac{1}{2}\right)}{\Gamma\left(\frac{k}{2}\right)} \, .
\end{equation}
Here the value $D_d \theta_E$, substituted as an argument of $\omega_p$, is the linear size of the vacuum Einstein ring. \\

Let us now discuss the properties of the lens equation (\ref{le-final}) with (\ref{norm-total-delf}) in different situations:

(i) In absence of plasma (ordinary gravitational lensing in vacuum), the deflection is
\begin{equation}
\alpha(|x|) = \frac{1}{|x|}  \, ,
\end{equation}
and the well-known point-mass lens equation is recovered \citep{GL2}:
\begin{equation} \label{le-point-y-x}
y = x - \frac{1}{x} \,  .
\end{equation}
It has two solutions:
\begin{equation} \label{x-pm}
x_\pm = \frac{1}{2} \left( y \pm \sqrt{y^2+4}  \right) .
\end{equation}
The solution $x_+>0$ is usually called as a primary, it is located on the same side of the lens as the source. The secondary image, $x_-<0$, is located on the opposite side of the lens, see Fig. \ref{fig-microlens}a.

(ii) Our formalism can also allow to study lensing in absence of gravity, namely, so called plasma lensing. Plasma lensing is a lensing by the compact distributions of plasma. In contrast to the converging achromatic behaviour of the vacuum gravitational lenses, the frequency-dependent lensing occurs from the refraction of electromagnetic rays by the plasma along an observer's line of sight and has the diverging character for spherically symmetric distributions with falling density. Such plasma lensing has been recently studied numerically by methods of gravitational lens formalism in series of works of \citet{Er-Rogers-2018, Er-Rogers-2019a, Er-Rogers-2019b}, see also \citet{Clegg-1998}. Note that in studies of interstellar scintillation the plasma column density is used instead of the volume density, see \citet{Rickett-1977}.

On the basis of the present approach, the normalized deflection angle for the plasma lensing is
\begin{equation}
\alpha(|x|) = - B(|x|)  \, ,
\end{equation}
and the lens equation has the following form:
\begin{equation}
y = x + B(|x|) \frac{x}{|x|} \, .
\end{equation}

(iii) In the general case of simultaneous presence of gravity and plasma, the lens equation (\ref{le-final}) with an overall delfection angle (\ref{norm-total-delf}) can be solved only numerically. Each solution $x_i$ of the lens equation gives us the image of the source.

Number of images depends on the plasma distribution and source position. This can be explained by plotting the right hand side of eq. (\ref{le-final}), with account of (\ref{norm-total-delf}), which we will denote as $F(x)$:
\begin{equation} \label{Fx-defin}
F(x) \equiv x - \alpha(|x|) \frac{x}{|x|} \,  .
\end{equation}
Every intersection of the horizontal line $y$ (source position) with the curve $F(x)$ determines a solution of the lens equation (\ref{le-final}). Thus, the number of images is determined by the number of these intersections.

\section{Numerical calculation of the microlensing light curve} \label{sec-num-magnif}

\begin{figure*}
\includegraphics[width=0.95\textwidth]{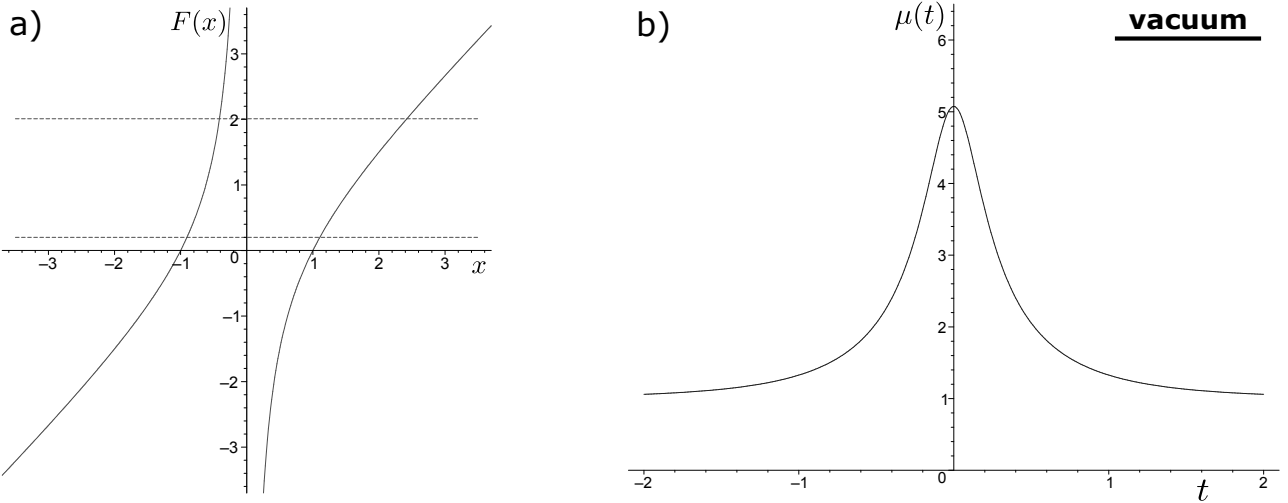}
\caption{Microlensing light curve for vacuum ($B_0=0$) with other parameters corresponding to Fig. \ref{fig-k3} and \ref{fig-k32}. See captions of those figures for more details. In the vacuum case, we always have two images (two intersections of horizontal line with function $F(x)$), which leads to the usual gravitational microlensing light curve.} 
\label{fig-vacuum}
\end{figure*}

In the axially symmetric case, the magnification factor (ratio of the flux from image, and from the unlensed source) can be written as \citep{GL1, GL2}
\begin{equation} \label{mu-i-def}
\mu_i =  \frac{x}{y} \frac{dx}{dy} \, .
\end{equation}

In the vacuum case of a point-mass lens, the magnification can be calculated fully analytically. The lens equation for the point-mass lens (\ref{le-point-y-x}) can be solved analytically, which gives explicit functions of positions of two images, $x_+(y)$ and $x_-(y)$, see (\ref{x-pm}). Substitution of $x(y)$ into (\ref{mu-i-def}) allows to write down magnifications of every image as an explicit function of $y$.

The total magnification is a sum of two magnitudes of individual images, and is equal to \citep{GL2}:
\begin{equation} \label{mu-tot-vacuum}
\mu = |\mu_1| + |\mu_2| = \frac{y^2+2}{y \sqrt{y^2+4}} \,  .
\end{equation}

The microlensing phenomenon is based on the change of the magnification with time due to the relative motion of the lens across the observer-source line-of-sight. The characteristic time scale of the magnification change is given by the Einstein time
\begin{equation}
t_E = \frac{D_d \theta_E}{v_\perp} \,  ,
\end{equation}
where $v_\perp$ is a transverse velocity of the lens relative to the source-observer line-of-sight \citep{GL2, Dodelson-GL, Keeton-book}. Since the magnification (\ref{mu-tot-vacuum}) depends on the source position and the motion is relative, microlensing is described by the source motion relative to the lens \citep{GL2}. Thus, during microlensing event, the source is moving and $y$ is changing with time, see Fig. \ref{fig-microlens}c. Usually the source trajectory is represented as \citep{GL2}:
\begin{equation}
y(t) = \sqrt{y_0^2 + \left(\frac{t-t_0}{t_E} \right)^2 }  \,  .
\end{equation}
Here $t_0$ is the time at which the source-lens separation $y(t)$ is minimal, $y(t_0)=y_0$. The value $y_0$ may be called as an impact parameter of the moving source (it should not be confused with the impact parameter $b$ of the light ray). For simplicity, hereinafter we will take $t_0=0$ and $t_E=1$, which is equivalent to measuring of time in the units of $t_E$. Substituting $y(t) = (y_0^2 + t^2)^{1/2}$ into (\ref{mu-tot-vacuum}), we obtain the microlensing light curve $\mu(t)$. In Fig. \ref{fig-microlens}e we plot the microlensing light curves for different values of the source impact parameter $y_0$.

In the case of plasma presence, the lens equation is more complicated and cannot be solved analytically. It means that we cannot obtain an explicit function $x(y)$ of each image. To calculate the magnification in this case, we rewrite eq.(\ref{mu-i-def}) as a function of $x$ only. With a help of the lens equation (\ref{le-final}), the magnification of each image takes form:
\begin{equation} \label{mu-i}
\mu^{-1}_i =  \left. \frac{y}{x} \frac{dy}{dx} \right|_{x=x_i} = \left.  \frac{F(x)}{x} \frac{dF(x)}{dx} \right|_{x=x_i}  \, .
\end{equation}

The advantage of this approach is that the analytical dependence $F(x)$ is known (see eq.(\ref{Fx-defin})), and the only value we need to calculate numerically is a solution of the lens equation, $x_i$. By formula (\ref{mu-i}), we obtain the magnification factor $\mu_i$ of each image as a function of its angular position $x_i$ found numerically. The total magnification is a sum of absolute values of all images magnifications:
\begin{equation} \label{mu-tot}
\mu = \sum_i |\mu_i|  \,  .
\end {equation}

For the reader convenience, we summarize the suggested method of the microlensing light curve calculation in a step-by-step procedure:

1. Choose the distribution of the plasma number density $N(r)$, calculate the angle of the refractive deflection as a function of $b$ by the formula (\ref{alpha-refr-b}), obtain the function $B(|x|)$ by the formula (\ref{B-defin}), and substitute it into (\ref{norm-total-delf}) and (\ref{le-final}). For power-law density profiles (\ref{N-power}), the function $B(|x|)$ is given by (\ref{B-power}) with the constant (\ref{B0}).

2. Fix some value of the impact parameter $y_0$ of the source and present $y$ as
\begin{equation} \label{y-sqrt}
y = \sqrt{y_0^2 + t^2}    \, ,
\end{equation}
where $t$ is the time in units of $t_E$.

3. Choose some interval of time. Then, for every $t$ from this interval, solve the lens equation (\ref{le-final}) numerically, which gives us a set of solutions $x_i$ (the angular positions of the images).

4. Calculate $\mu_i$ for every $x_i$ by formula (\ref{mu-i}).

5. Calculate $\mu$ by formula (\ref{mu-tot}) and obtain the magnification curve $\mu(t)$.

\section{Microlensing light curves for power-law density profile with the index 3} \label{sec-k3}

\begin{figure*}
\includegraphics[width=0.9\textwidth]{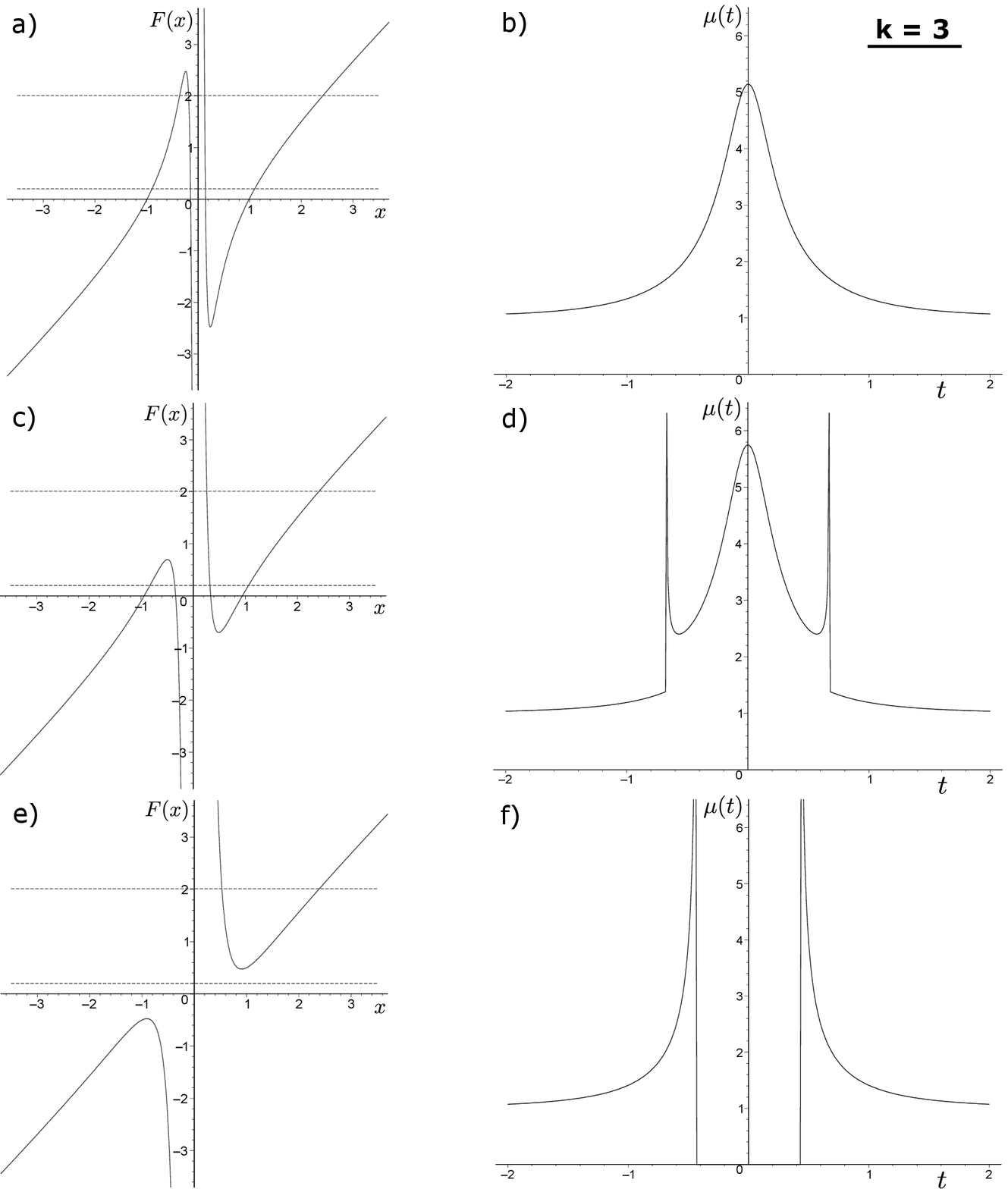}
\caption{Microlensing light curves in the case of a power-law distribution with $k=3$, see eqs (\ref{k3-N}), (\ref{k3-B}), (\ref{k3-le}), plotted for different values of $B_0$ which characterizes the plasma influence and is defined by (\ref{B0}). At the left column we plot the right hand side of lens equation (\ref{le-final}), namely the function $F(x)$ from (\ref{Fx-defin}), for some value of $B_0$. At the right column we plot the microlensing light curve $\mu(t)$ for the same value of $B_0$ using dependence $y=(y_0^2 + t^2)^{1/2}$ with $y_0=0.2$ and $-2 \le t \le 2$, see formula (\ref{y-sqrt}) and end of previous Section for details. The upper horizontal dashed line at the left figures is the value of $y$ for $t = \pm 2$, on the lower horizontal line $y=y_0$, corresponding to $t=0$. During the microlensing event, the value $y$ changes from the upper to lower line, and then back. a,b) $B_0 = 0.02$ (Influence of plasma is small.) For this value of $B_0$, there are always four images in the plotted range of $t$. In the plotted range of $t$ the microlensing light curve does not differ much from the vacuum case; the maximum of the curve becomes slightly higher. c,d) $B_0 = 0.1$. For this value, the change of number of images ($4 \leftrightarrow 2$) happens during the change of $y$, which leads to noticeable change in the curve $\mu(t)$. Similar spikes also happen outside the plotted range of $t$ at picture b). e,f) $B_0 = 0.5$. For this value of $B_0$, the refraction is strong enough  to make the images invisible for observer: for small enough values of $y$ there are no solutions of the lens equation and, correspondingly, there are no images of the source. Therefore we have a 'hole' in the center of microlensing light curve. High peaks in magnification curves (d,f) correspond to source crossing the caustics when the magnification becomes infinite and new pair of images appears or disappears \citep{GL2, Keeton-book, Alex-2011}, see Fig. \ref{fig-caustics}. If a simulation uses not a point source, but a source of a finite size, high peaks are smoothed.} 
\label{fig-k3}
\end{figure*}

\begin{figure*}
\begin{center}
\includegraphics[width=0.80\textwidth]{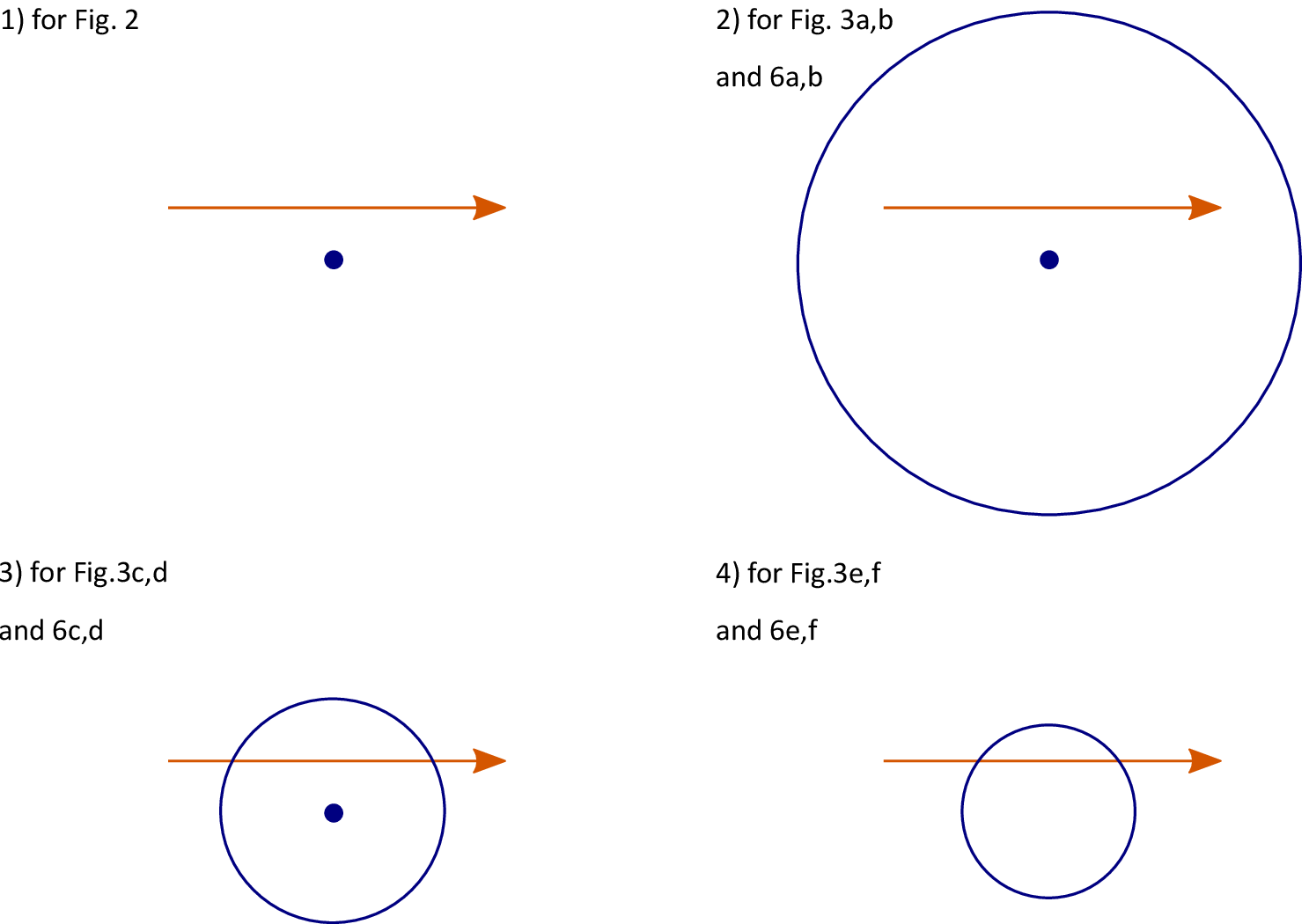}
\end{center}
\caption{Caustic structure in the source plane corresponding to situations shown at Figs.\ref{fig-vacuum}, \ref{fig-k3}, \ref{fig-k32}. Horizontal arrow shows the source trajectory; all scales are presented conventionally. Axisymmetric lenses produce caustics that are points or circles \citep{Keeton-book}. The number of images of the given lens increases or decreases by two when the source crosses the caustic. The magnification becomes infinite when the right hand side of (\ref{mu-i}) equals to zero.
It happens if either (or both) of the factors vanishes, i.e., $F(x)=0$ (tangential caustic) or $dF/dx=0$ (radial caustic). Presence or disappearance of solutions of these two equations allow us to describe qualitatively the behaviour of microlensing light curves. 1) Vacuum case ($B_0=0$), and $F(x) = x - 1/x$. We have only one possibility to obtain zero: when $F(x)=0$, which means $y=0$ with $x=1$. It gives us one caustic point in the source plane (bold point at the pictures). Since the trajectory is near caustic, we have the high (finite) maximum in the center of microlensing light curve. 2) When plasma is added, the function $F(x)$ becomes more complicated, and there arises a possibility to have $dF/dx=0$ (radial caustic), which gives us the caustic circle (solid circle at the pictures). If the strength of plasma influence is small, the circle is big. In this case, the trajectory of source (for given range of $t$) is completely inside the caustic circle, so we don't see spikes at the corresponding pictures. For bigger range of $t$, we would see spikes when the source is crossing the caustic circle. 3) For bigger strength of plasma influence (bigger plasma density or smaller photon frequencies), the caustic circle becomes smaller. Now the chosen trajectory of source is crossing this caustic, and we have spikes at corresponding light curves, together with central maximum due to the central caustic point. 4) The plasma influence is increasing with increase of its density, leading to shrinking of the radial caustic, until its merging with the degenerate tangential caustic point at the centre. After further density increasing the caustic re-expands, with no images formed for sources inside the caustic. As a result, the hole is formed. In this case eq.$dF/dx=0$ have solution but solution of eq.$F(x)=0$ disappears, so here we have the caustic circle but don't have the caustic point in the center.}
\label{fig-caustics}
\end{figure*}

\begin{figure*}
\begin{center}
\includegraphics[width=0.95\textwidth]{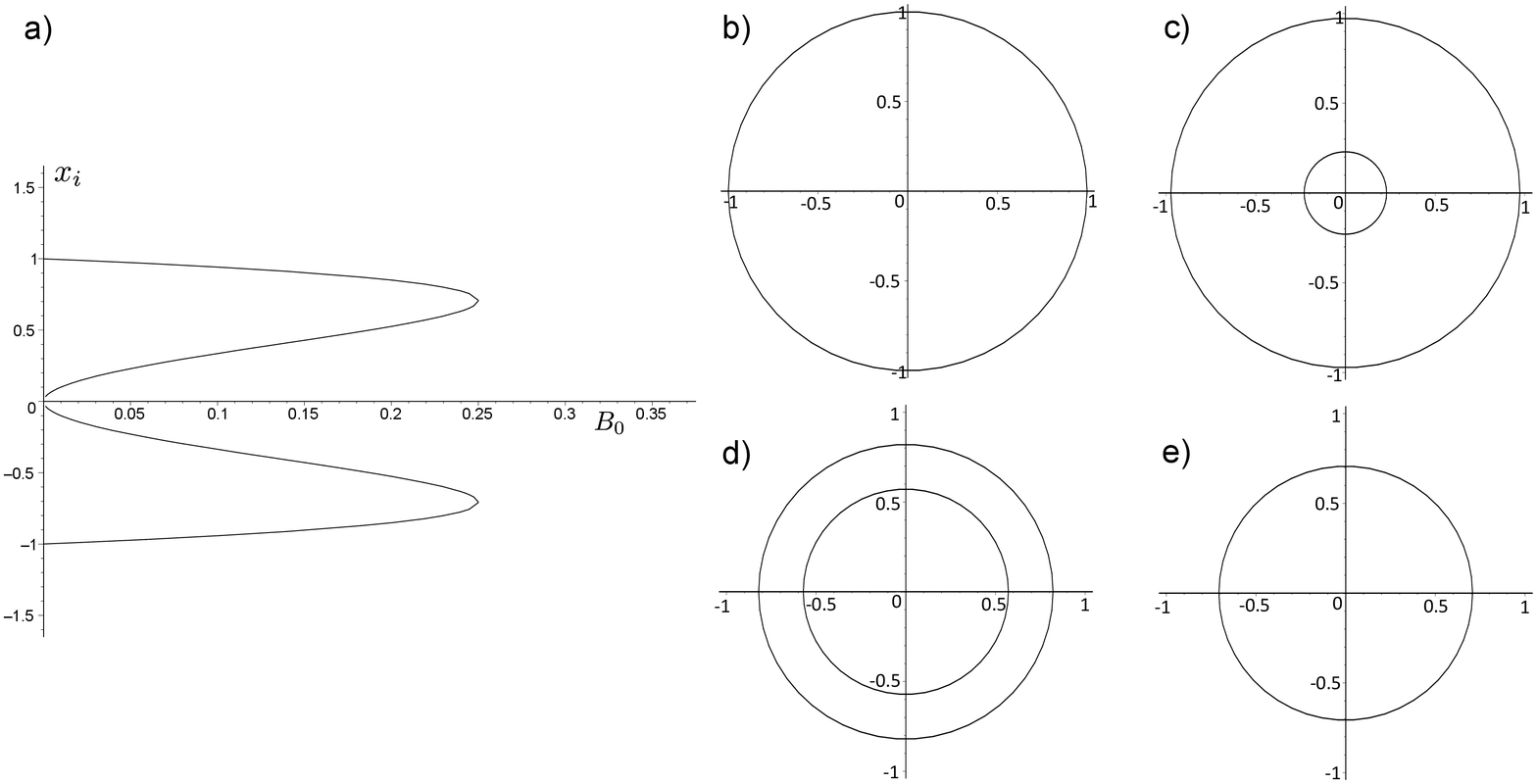}
\end{center}
\caption{Ring-images in the case of a gravitational lensing in plasma, see also Fig. 2.19 in \citet{Bliokh-Minakov-1989}. a) The dependence of four solutions $x_i$ of equation (\ref{k3-le-y-0}) as functions of $B_0$. b) $B_0=0$. Einstein ring in vacuum. c) $B_0=0.05$. Two rings, the bigger one is smaller than vacuum Einstein ring. d) $B_0=0.22$. With increasing $B_0$, the large ring becomes smaller, and the small ring becomes larger. e) $B_0=0.25$. Two rings merge into one.  For $B_0>0.25$ there are no images. }
\label{fig-rings}
\end{figure*}

Let us start from the case of the power-law density profile with $k=3$. This case is interesting because it allows analytical analysis of the lens equation (see the end of this section).

In this case, the plasma number density is:
\begin{equation} \label{k3-N}
N(r) = N_0 \left( \frac{r_0}{r} \right)^3  \,  .
\end{equation}
Correspondingly, the function $B(|x|)$ has a form:
\begin{equation} \label{k3-B}
B(|x|) = \frac{B_0}{|x|^3}  \,  ,
\end{equation}
and the lens equation can be presented as
\begin{equation}
y = x - \left( \frac{1}{|x|} - \frac{B_0}{|x|^3} \right) \frac{x}{|x|} \, ,
\end{equation}
or
\begin{equation} \label{k3-le}
y = x - \frac{1}{x} + \frac{B_0}{x^3}  \,  .
\end{equation}
Results of the numerical calculation of microlensing events are shown in Fig. \ref{fig-vacuum} ($B_0=0$, vacuum) and Fig. \ref{fig-k3} ($B_0 \ne 0$), where we plot the function $F(x)$ (left column) and corresponding microlensing light curve (right column). The number of images is determined by the number of intersections of the horizontal line $y$ with the curve $F(x)$. We choose a range of interval of $t$ from $-2$ to $2$, and during the microlensing event the value $y$ changes from the maximum (the upper horizontal line at left figures, which corresponds to $t=-2$) value to the minimum value (the lower horizontal line, which corresponds to $t=0$), and then back to the upper horizontal line at $t=2$. See caption of Fig. \ref{fig-k3} for more details.

The vacuum case (Fig.\ref{fig-vacuum}a,b) corresponds to $B_0=0$, and for this case we always have two intersections of the horizontal line $y$ with the function $F(x)$. Therefore we always have two images.

For small $B_0$ (Fig.\ref{fig-k3}a,b) a new pair of images appears with small $x$ (close to the lens). Therefore in this case we have four images. (We discuss mainly the range $y \lesssim 1$.) But the curve itself differs not much from the vacuum case (in the presented range of $t$).

For larger values of $B_0$ (Fig.\ref{fig-k3}c,d) the change of the number of images ($4 \to 2$) happens in the range $y \lesssim 1$, and the magnification curve noticeably changes.

For even larger values of $B_0$ (Fig.\ref{fig-k3}e,f) there is a region when there is no intersections of $y$ with $F(x)$, and there are no solutions of the lens equation. Therefore, for some time during microlensing event, the image is absent. This means that the diverging refractive deflection of the rays is becoming so strong that no ray from the source can reach the observer.
In this case, in the center of the microlensing light curve we have a `hole' (instead of a 'hill'). 
The 'hole' of the same origin was found in the paper of \citet{Clegg-1998} (see Figures 2,3,4,5 of that paper) and recent paper of \citet{Er-Rogers-2018} (see Figures 3,5,7 of that paper) where they investigated the diverging refractive lensing by the compact distributions of plasma (in absence of the gravitational lens).\\

For $k=3$, the formation of new images due to plasma presence and the complete disappearance of images in the case of a large refraction can be solved analytically for the particular case at $y=0$, see the discussion in \citet{Bliokh-Minakov-1989}.

For $y=0$ (perfect alignment of the source, lens and observer), Eq. (\ref{k3-le}) reduces to
\begin{equation} \label{k3-le-y-0}
0 = x - \frac{1}{x} + \frac{B_0}{x^3}  \,  ,
\end{equation}
and can be solved fully analytically. For $B_0=0$ (vacuum) we have two solutions of different signs and of the same modulus, $x=\pm 1$, it is a vacuum Einstein ring. For $0 < B_0 < 1/4$ we have four solutions:
\begin{equation}
x_{1,2} = \pm \frac{1}{2} \sqrt{2 + 2 \sqrt{1 - 4B_0}} \quad \text{(first ring)} \, ,
\end{equation}
\begin{equation}
x_{3,4} = \pm \frac{1}{2} \sqrt{2 - 2 \sqrt{1 - 4B_0}} \quad \text{(second ring)} \, .
\end{equation}
For $B_0=1/4$ four solutions merge into two solutions:
\begin{equation}
x_{1,2} = \pm \frac{\sqrt{2}}{2}  \, ,
\end{equation}
so we have one ring. For $B_0>1/4$ there are no solutions. It means that diverging refraction is so large that the images becomes invisible for observer. In Fig. \ref{fig-rings} we present properties of these solutions.

\section{Microlensing light curves for power law density profile with the index 3/2} \label{sec-k32}

\begin{figure*}
\begin{center}
\includegraphics[width=0.95\textwidth]{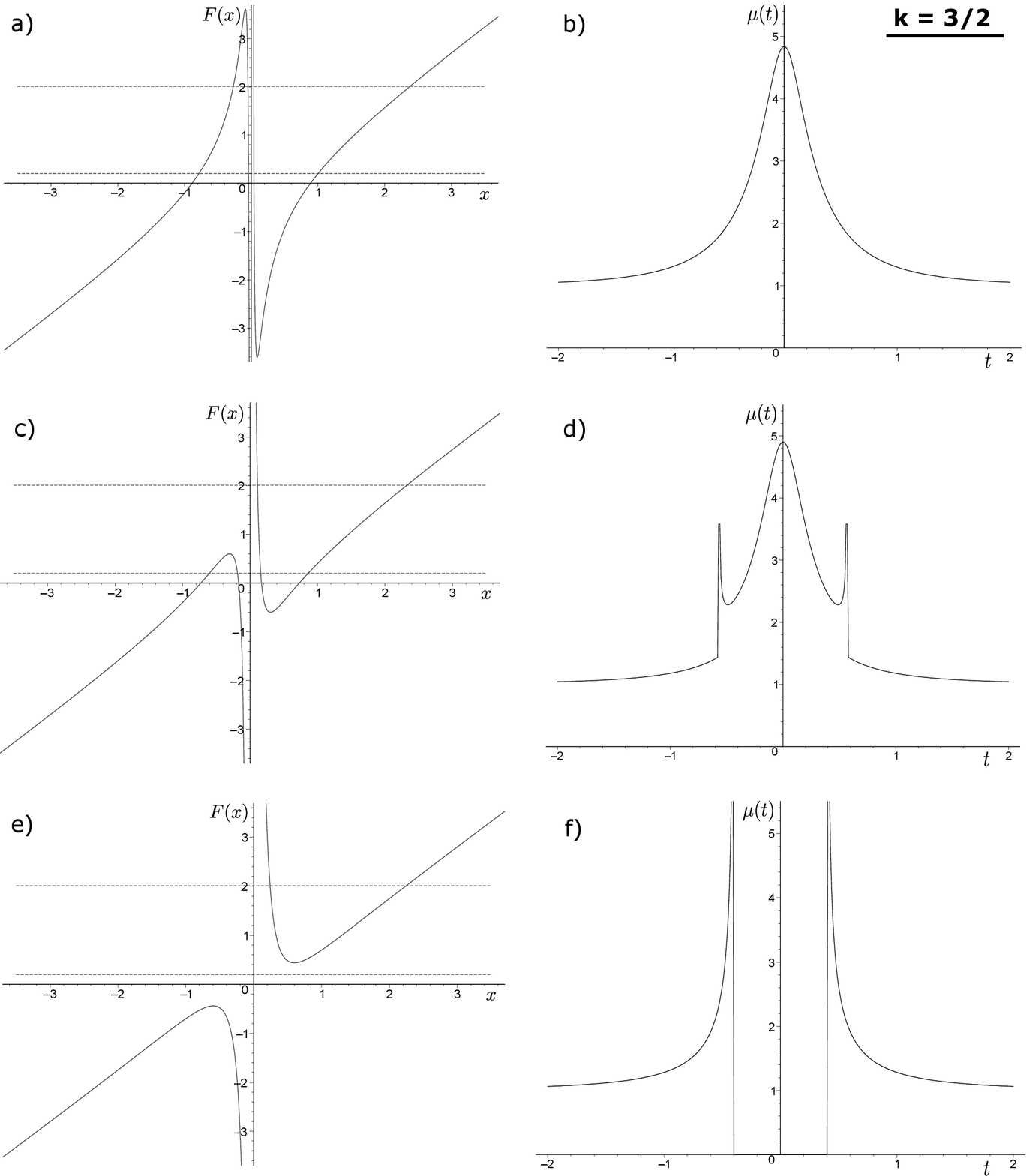}
\end{center}
\caption{Microlensing light curves in the case of a power-law distribution with $k=3/2$, see eqs (\ref{k32-N}), (\ref{k32-B}), (\ref{k32-le}), plotted for different values of $B_0$; for more details see the caption of the  figure 2. a,b) $B_0 = 0.2$. For this value of $B_0$, there are always four images in the plotted range of $t$. In the plotted range of $t$ the microlensing light curve does not differ much from vacuum case; the maximum of curve becomes slightly smaller. c,d) $B_0 = 0.4$. For this value, the change of number of images ($4 \leftrightarrow 2$) happens during the change of $y$, which leads to noticeable change in the curve. e,f) $B_0 = 0.7$. For this value of $B_0$, the refraction is strong enough to make the images invisible for observer. Therefore we have a hole in the center of microlensing light curve.
 }
\label{fig-k32}
\end{figure*}

The power-law density distribution with $k=3/2$ is physically motivated, since it corresponds to spherically symmetric accretion of free falling particles (dust), see Section VI in paper \citet{Perlick-Tsupko-BK-2015}.

The plasma number density is defined as
\begin{equation} \label{k32-N}
N(r) = N_0 \left( \frac{r_0}{r} \right)^{3/2}  \,  .
\end{equation}
The function $B(|x|)$ has a form:
\begin{equation} \label{k32-B}
B(|x|) = \frac{B_0}{|x|^{3/2}}  \,  ,
\end{equation}
and the lens equation can be presented as
\begin{equation} \label{k32-le}
y = x - \left( \frac{1}{|x|} - \frac{B_0}{|x|^{3/2}} \right) \frac{x}{|x|}  \,  .
\end{equation}

Microlensing light curves $\mu(t)$ calculated with this density profile and different values of $B_0$ are shown in Fig. \ref{fig-k32}. In general, all effects are the same as in Fig. \ref{fig-k3}.\\

\section{Observational prospects} \label{sec-observ}

In this Section, we discuss the observational prospects. We introduce the criterium of presence of 'hill-hole' effect and discuss the applications to observations.

As discussed in previous studies \citep{BK-Tsupko-2009, BK-Tsupko-2010, Er-Mao-2014, BKTs-Universe-2017}, plasma effects in the strong lens systems are significant mainly for the radio band. It is connected with relatively small plasma density in areas where the light propagates. Speaking about strong lens systems, plasma presence can provide only a small change in positions and magnifications of images in comparison with the vacuum case.

Here, in case of microlensing, the situation is more favorable. First of all, the gravitational deflection itself is smaller by several orders of magnitude. Therefore, the refractive deflection of a given magnitude will play a much larger role in microlensing than in strong lens systems. Second, in the microlensing situations, the refraction can be quite large due to high concentration of plasma: when a light ray travels through the plasma atmosphere of a star or through the accretion disk around a black hole. In this case, the light trajectory is determined by the simultaneous action of both gravity and plasma. The refraction can be large enough to do a significant change in the physical situation: the images can completely dissapear (become invisible).

\subsection{Condition on $B_0$ value for the 'hole' formation} \label{subsec-B0}

To make the microlensing images invisible and to see a hole instead of a hill on the light curve, the refractive deflection should compensate or overcome the gravitational deflection. Sufficient condition, as can be readily seen in the left panels of Figures \ref{fig-k3} and \ref{fig-k32}, is that the function $F(x)$ at local minimum $x_1$ (at $x>0$) is positive, so that there is a gap on the vertical axis that does not correspond to any value on the horizontal axis \footnote{We thank the anonymous Referee for indication of this point and also for suggestion of calculation of the condition for the hole formation on basis of caustic analysis (see below in the text)}.

Let us consider the case $k=3$, where analytical analysis is possible. In this case, the function $F(x)$ is written as
\begin{equation}
F(x) = x - \frac{1}{x} + \frac{B_0}{x^3} = \frac{1}{x^3} (x^4 - x^2 + B_0)  \,  .
\end{equation}
From analysis of polynomial $(x^4 - x^2 + B_0)$ it is easy to see that $F(x_1)>0$ for $B_0>1/4$. Obviously, it agrees with the results presented at the end of Section \ref{sec-k3}.

Exact condition for the hole formation can be also found for arbitrary $k>1$. Let us write $F(x)$ for arbitrary value of $k$ and for positive $x$:
\begin{equation} \label{arb-k-0}
F(x) = x -\frac{1}{x} + \frac{B_0}{x^k} \, .    
\end{equation}
Local minimum $x_1$ of function $F(x)$ can be found from the solution of eq. $dF/dx=0$ or, explicitly:
\begin{equation} \label{arb-k-1}
0 = 1 + \frac{1}{x^2} - k \frac{B_0}{x^{k+1}}    \, .
\end{equation}
For the hole formation we need to have $F(x_1)>0$, so the boundary value of $B_0$ corresponds to the equation $F(x)=0$ or, explicitly:
\begin{equation} \label{arb-k-2}
0 = x -\frac{1}{x} + \frac{B_0}{x^k} \, .    
\end{equation}
Solving equations (\ref{arb-k-1}) and (\ref{arb-k-2}) together, we obtain 
\begin{equation}
x_1 = \left( \frac{k-1}{k+1}    \right)^{1/2}
\end{equation}
and the boundary value of $B_0$:
\begin{equation} \label{arb-k-4}
B_0 = 2 \frac{(k-1)^{\frac{k-1}{2}}}{(k+1)^{\frac{k+1}{2}}}  \, .
\end{equation} 
In particular, for $k=3/2$ it gives $B_0 \simeq 0.535$, which agrees with the numerical results presented at Fig. \ref{fig-k32}.

The expression (\ref{arb-k-4}) can be also found from analysis of behaviour of caustics (critical curves) presented at Fig. \ref{fig-caustics}. The hole is formed when merger of caustics happens. The conditions on the radial and tangential caustics (critical curves) are $dF/dx=0$ and $F(x)=0$. Therefore, from condition of merging we obtain the system of the same two equations (\ref{arb-k-1}) and (\ref{arb-k-2}), and the same expression (\ref{arb-k-4}).

By order of magnitude, we may formulate the following criterium for the hole formation as:
\begin{equation} \label{B0-criterium}
B_0>1 \, .
\end{equation}
In particular, it agrees also with numerical results for $k=3/2$ (Fig. \ref{fig-k32}).

If $B_0 \gg 1$, the source is not visible (eclipse). The most interesting in observations would be to see the characteristic light curves similar to examples shown in d) and f) of Figures \ref{fig-k3} and \ref{fig-k32}. We can formulate the expectations for different values of $B_0$:

$B_0 \ll 1$ -- light curves are similar to vacuum one;

$B_0 \sim 1$ -- characteristic light curves as in d) and f) of Figures \ref{fig-k3} and \ref{fig-k32}.\\

Let us find out the physical sense of (\ref{B0-criterium}). Considering the explicit expression for $B_0$ from (\ref{B0-2}), assuming $D_{ds}$ and $D_s$ are of the same order, and neglecting non-dimensional numerical coefficients ($\pi$, $\Gamma(k)$ etc), we obtain the following physical condition for the hole formation:
\begin{equation} \label{eq-phys-crit}
\frac{\omega_p^2(D_d \theta_E)}{\omega_0^2} > \theta_E \,  , \quad \mbox{or} \quad  \left| \alpha_{refr}(D_d \theta_E) \right|> \theta_E \, .
\end{equation}
Here the ratio of frequencies is of the order of the refractive deflection, $\alpha_{refr}(b) \sim \omega_p^2(b)/\omega_0^2$, and $D_d \theta_E$ is the linear size of vacuum Einstein radius. By the order of magnitude (assuming that $D_{ds}$ and $D_s$ are of the same order), $\theta_E$ can be considered as the Einstein deflection $4m/b$ with $b=D_d \theta_E$ substituted: $\theta_E \sim \alpha_{grav} (D_d \theta_E)$. Therefore, we can rewrite the condition of the hole formation as
\begin{equation} \label{eq-phys-crit}
\left| \alpha_{refr} (D_d \theta_E) \right|> \alpha_{grav} (D_d \theta_E) \, .
\end{equation}

Using $\lambda = 2\pi c/ \omega_0$ (here we neglect the difference of light velocity in the  medium from the vacuum), we can write the following numerical value:
\begin{equation} \label{est-01}
B_0 \sim \frac{\omega_p^2}{\omega_0^2 \theta_E} = 0.9 \cdot 10^{-13} N \lambda^2 \frac{1}{\theta_E} \, ,
\end{equation}
where $N$ is in cm$^{-3}$, $\lambda$ in cm and $\theta_E$ in radians.

\subsection{Quasar microlensing}

Two possible scenarios of applications of hill-hole effect in observations can be proposed.\footnote{We are thankful to Xinzhong Er for useful suggestions.} 
First is a quasar microlensing (stars of lensing galaxy act as lenses for multiple macroimages of quasar). For this case, a typical angular size is \citep{GL2, Wambsganss-2010}
\begin{equation}
\theta_E \sim 10^{-6} \left( \frac{M}{M_\odot} \right)^{1/2} \; \mbox{arcsec} \, . 
\end{equation}
Quasar microlensing light curves are now being obtained in all parts of the electromagnetic spectrum from the radio through the infrared and optical range to the X-ray regime \citep{Wambsganss-2010}. The numerical value of $B_0$ for this case is
\begin{equation}
B_0^{(q)} \simeq  0.02 \left( \frac{N}{1 \, \mbox{cm}^{-3}} \right) \left( \frac{\lambda}{1 \, \mbox{cm}} \right)^2   \left( \frac{M_\odot}{M} \right)^{1/2}   \, .
\end{equation}

In the radio band, the hill-hole effect should be observable even without high concentration of plasma around the lens: interstellar number density inside galaxy is already big enough to lead to significant influence. For radio frequency $\nu = 375$ MHz ($\omega_0 = 2 \pi \nu$ and $\lambda \simeq 80$ cm) and interstellar number density inside galaxy $N=10$ cm$^{-3}$ \citep{Er-Mao-2014} we obtain: $B_0 \simeq 10^3$. We can conlcude that the plasma effects may be of crucial importance for quasar microlensing in the radio band. For radio observations of lensed quasars see, e.g., \citet{radio-1994, radio-1996, radio-2015-R, radio-2015-J, Borra-2014}. We note that effects of scattering by inhomogeneities of the turbulent plasma in galaxy may wash away the effect in observations. At the same time, since microlensing observations in the radio range are carried out, we can expect that such effects manifest themselves not in all microlensing events.

Plasma may be important also for so called 'chromatic microlensing' \citep{Kayser-1986, Wambsganss-1991, Eigenbrod-2011}. Whereas gravitational lensing in vacuum is achromatic, the magnification light curves can be chromatic if the surface brightness profile of the source is different at different wavelengths. It can be described as 'achromatic microlensing of chromatic source'. In presence of plasma, the deflection itself becomes chromatic, and it should be taken into account during modelling.

\subsection{Galactic microlensing}

Second application is a Galactic microlensing (made by star in our Galaxy). In the case of Galactic microlensing, we are interested mainly in optical and infrared bands. The planned project WFIRST \citep{WFIRST-1, WFIRST-2} will extend the current sensitivity of the microlensing method down to masses of about 0.1 of the Earth's mass \footnote{https://wfirst.gsfc.nasa.gov/exoplanets\_microlensing.html}. Observations will be performed in optical and near-infrared bands.

Typical angular size $\theta_E$ for Galactic microlensing is \citep{GL1, Mao-review}:
\begin{equation}
\theta_E \sim 10^{-3} \left( \frac{M}{M_\odot} \right)^{1/2}  \; \mbox{arcsec} \, .   
\end{equation}
For such value of $\theta_E$ we obtain the numerical value of $B_0$ as
\begin{equation}
B_0^{(g)} \simeq  2 \cdot 10^{-13} \left( \frac{N}{1 \, \mbox{cm}^{-3}} \right) \left( \frac{\lambda}{1 \, \mbox{mkm}} \right)^2   \left( \frac{M_\odot}{M} \right)^{1/2}   \, .
\end{equation}

Speaking about infrared frequencies, with $\lambda = 1$ mkm, and lenses of Solar mass, we obtain that we need $N> 10^{13}$ cm$^{-3}$ to get $B_0 >2$. Such a high value of plasma density is reachable near stars and compact objects. In accretion disc around a black hole in binary system the mass density may reach the values of about $10^{-5} \; \mbox{g/cm}^3$ \citep{Artemova-2006}. The same order of magnitude of the mass density can take place in levitating athmosphere of compact objects \citep{Rogers-2017b, Wielgus-2016}. It leads to the number density of the order of $N \sim 10^{19} \, \mbox{cm}^{-3}$. Although this number is very high, we note that the light ray is travelling at large distances from lens, with the impact parameter about 1 AU. Therefore, light ray may not propagate through the regions of high density, and possibility of observation of effect in the case of isolated black hole or star is questionable.

Apparently, the most promising candidate is a binary system consisting of an ordinary star and a compact object (or star with planet, possibly with propotoplanetary disc). To obtain the effect, the light ray experiencing the gravitational influence of both components should pass near one of them where the plasma density is high. Numerical modelling of binary system with some plasma distributions around components is required for detail investigation of this problem.

\section{Influence of other physical effects in dense plasma} \label{sec-Th-FF}

To observe the hill-hole effect, the plasma medium should be optically thin. In the case of optically thick medium, images will disappear due to scattering and absorption. Here we discuss the conditions for the optical depth with respect to different physical phenomena.

\subsection{Thomson scattering}

For the rays to reach the observer, plasma should be optically thin with respect to the Thomson scattering. It means that 
Thomson optical depth $\tau$ should be less than unity,
\begin{equation} \label{th-00}
\tau < 1 \, .
\end{equation}

Let us calculate the Thomson optical depth for spherically symmetric power-law plasma distribution $N(r)$ from (\ref{N-power}). We consider the light ray whose unperturbed trajectory is a straight line parallel to the axis $z$ with the impact parameter $b$. Thomson optical depth $\tau$ for such trajectory is calculated as:
\begin{equation}
\tau = \int \limits_{-\infty}^{\infty} \sigma_T N(r) \, dz = 2 \int \limits_0^{\infty} \sigma_T N(r) \, dz =
\end{equation}
\begin{equation}
= 2 \sigma_T N_0 r_0^k \int \limits_0^{\infty} \frac{dz}{(z^2+b^2)^{k/2}} =
\end{equation}
\begin{equation}
= \sigma_T N_0  \, \frac{r_0^k}{b^{k-1}} \, \frac{\sqrt{\pi} \Gamma(k/2-1/2)}{\Gamma(k/2)} \, .
\end{equation}
Here $\sigma_T$ is the Thomson cross section. Neglecting non-dimensional numerical coefficients, we can write:
\begin{equation}
\tau \simeq \sigma_T N(b) \, b  \, , \;\; \mbox{where} \;\; N(b) = N_0 \left( \frac{r_0}{b} \right)^k  \,  .
\end{equation}

For further discussion, let us write the Thomson optical depth by order of magnitude as
\begin{equation}
\tau \simeq \sigma_T N L \, ,
\end{equation}
where $L$ is characteristic size of the region of high plasma concentration with characteristic value $N$ of number density. Now we write the condition (\ref{th-00}) as
\begin{equation}
\sigma_T N L < 1 \, ,
\end{equation}
or, numerically, as
\begin{equation}
10^{-11} \left(  \frac{N}{1 \, \mbox{cm}^{-3}}  \right) \left(  \frac{L}{1 \, \mbox{AU}}  \right)   < 1   \, .
\end{equation}
We note that, if we consider not a spherically symmetric situation, but, for example, the accretion disk, then the characteristic length will be several orders of magnitude less.

\subsection{Thermal Bremsstrahlung (free-free) absorption}

Absorption coefficient $\alpha^{ff}_\nu$ is written (for $h \nu \ll kT$) as \citep{Ryb-L-2004}
\begin{equation}
\alpha^{ff}_\nu = \frac{4e^6}{3m_e kc} \left( \frac{2 \pi}{3k m_e} \right)^{1/2} T^{-3/2} Z^2 N_e N_i \nu^{-2} \bar{g}_{ff}  \, ,
\end{equation}
or, numerically,
\begin{equation}
\alpha^{ff}_\nu = 0.018 \, T^{-3/2} Z^2 N_e N_i \nu^{-2} \bar{g}_{ff} \; \mbox{cm}^{-1} \, .
\end{equation}
Here $Ze$ is ion charge, $N_i$ is ion number density (in cm$^{-3}$), $N_e$ is electron number density,  $\bar{g}_{ff}(T,\nu)$ is velocity averaged Gaunt factor (by order of magnitude it can be set to unity), $T$ is in K, $\nu$ is in Hz.

Optical depth $\tau$ should be less than unity. Further, we take $N_i=N_e=N$, $Z=1$, $\bar{g}_{ff}=1$ and obtain:
\begin{equation}
\alpha^{ff}_\nu = 0.018 \, T^{-3/2}  N^2 \nu^{-2}  \; \mbox{cm}^{-1} \, .
\end{equation}
For the light ray moving along axis $z$ with impact parameter $b$ in plasma distribution $N(r)$ from (\ref{N-power}), the optical depth for absorption is calculated as
\begin{equation}
\tau = \int \limits_{-\infty}^{\infty} 0.018 \, T^{-3/2}  N^2(r) \, \nu^{-2} \, dz =
\end{equation}
\[
= 0.018 \, N^2(b) \, b \, T^{-3/2} \nu^{-2}  \frac{\sqrt{\pi} \Gamma(k-1/2)}{\Gamma(k)} \, .
\]

By order of magnitude, we can write the condition $\tau<1$ as
\begin{equation}
0.018 \, \frac{N^2 L}{T^{3/2} \nu^2}   < 1 \, ,
\end{equation}
or, numerically, using $\nu=c/\lambda$, as
\begin{equation}
10^{-14} \left(  \frac{N}{1 \, \mbox{cm}^{-3}}  \right)^2   \left( \frac{10^3 \; \mbox{K}}{T}  \right)^{3/2} \left( \frac{\lambda}{1 \; \mbox{cm}}  \right)^2  \left(  \frac{L}{1 \, \mbox{AU}}  \right)    <   1   \, .
\end{equation}

\subsection{Cut-off plasma frequency}

For the wave to propagate in plasma, the photon frequency should be bigger than the plasma frequency, $\omega_0 > \omega_p$. 
Numerically, we obtain the condition:
\begin{equation} \label{Pl-01}
\omega_0 > 6 \cdot 10^4 \sqrt{N} \; \mbox{s}^{-1}   \,  ,
\end{equation}
or
\begin{equation} \label{Pl-02}
\lambda < 0.3 \cdot 10^7 \frac{1}{\sqrt{N}} \;  \mbox{cm}  \,  ,
\end{equation}
where $N$ is measured in cm$^{-3}$.

\section{Conclusions} \label{sec-conclusions}

(i) The influence of the plasma surrounding the lens on the effect of microlensing is investigated. In the presence of plasma around the lens, the deflection angle is determined by both the gravitational field of the lens and the chromatic refraction in the inhomogeneous plasma.

(ii) Lens equation for lensing in presence of gravity and non-homogeneous plasma is presented. In our approximations, both gravitational and refractive deflections are small, but the plasma deflection is not assumed to be much smaller than the gravitational one.

(iii) Microlensing light curves are calculated numerically for point-mass lens and two power-law density distributions of plasma, namely for $k=3$ and $k=3/2$. A variety of possible curves is found, depending on the plasma density and frequency of observations. Plasma influence is characterized by the value of $B_0$ from (\ref{B0}), (\ref{B0-2}), see discussion in Subsection \ref{subsec-B0}.

(iv) For small magnitudes of plasma influence, the microlensing light curve does not differ much from vacuum case. The maximum of curve in presence of plasma may be both higher and lower in comparison with vacuum case, depending on the plasma density distribution, see image b) at Figs \ref{fig-vacuum}, \ref{fig-k3}, \ref{fig-k32}.

(v) In the case of significant influence of plasma, the shape of the microlensing light curve is strongly  deformed in comparison with the vacuum case, see d) and f) of Figures \ref{fig-k3} and \ref{fig-k32}. If the refractive deflection is large enough to compensate or overcome the gravitational deflection, microlensing images can completely disappear for the observer due to divergence of rays. In this case, the 'hole' instead of the 'hill' is formed in the center of microlensing light curve. The search of such characteristic light curves can be an interesting target for future observations.

(vi) Observational prospects of the 'hill-hole' effect for quasar microlensing and for galactic microlensing are discussed (Section \ref{sec-observ}), as well as other effects in dense plasma (Section \ref{sec-Th-FF}).

\section*{Acknowledgments}

We thank the anonymous Referee for important comments. We are thankful to Dr Alexander Polnarev for useful and pleasant discussions. We are thankful also to Prof. Matthew Penny for discussion of observational prospects.

The reported study was partially supported by Russian Foundation for Basic Research (RFBR), project number 17-02-00760.












\bsp	
\label{lastpage}
\end{document}